\documentclass[aps,prl,twocolumn,groupedaddress]{revtex4-1}
\usepackage{graphicx}  
\usepackage{dcolumn}   
\usepackage{bm}        
\usepackage{amssymb}   
\usepackage{amsmath}
\usepackage{xfrac}

\begin{document}


\title{Three-dimensional particle-in-cell modeling of parametric instabilities 
near the quarter-critical density in plasmas}


\author{H. Wen$^{1,2}$}
\author{A. V. Maximov$^{1,2}$}
\author{R. Yan$^{1,2}$}
\altaffiliation[Now at ]{University of Science and Technology of China, Hefei, Anhui 230026, China}
\author{J. Li$^{1,2}$}
\altaffiliation[Now at ]{University of California, San Diego, CA 92093, USA}
\author{C. Ren$^{1,2,3}$}
\author{F. S. Tsung$^4$}
\affiliation{$^1$ Laboratory for Laser Energetics, University of Rochester, Rochester, New York 14623-1299, USA}
\affiliation{$^2$ Department of Mechanical Engineering, University of Rochester, Rochester, New York 14627, USA}
\affiliation{$^3$ Department of Physics and Astronomy, University of Rochester, Rochester, New York 14627, USA}
\affiliation{$^4$ Department of Physics and Astronomy, University of California Los Angeles, Los Angeles, California 90095, USA}


\date{\today}

\begin{abstract}
The nonlinear regime of laser-plasma interaction
including both
two-plasmon--decay (TPD) and stimulated Raman 
scattering (SRS) instabilities has been studied in three-dimensional (3-D) 
particle-in-cell simulations with parameters 
relevant to the inertial confinement fusion (ICF) experiments. 
SRS and TPD develop in the same region in plasmas, and the generation 
of fast electrons can be described accurately with only the full model 
including both SRS and TPD. The growth of instabilities in the linear stage 
is found to be in good agreement with analytical theories. 
In the saturation stage 
the enhanced low-frequency density perturbations
driven by the daughter 
waves of the SRS sidescattering can saturate the TPD and consequently inhibit 
the fast-electron generation. 
The fast-electron flux in 3-D modeling is up to an order of magnitude smaller 
than previously reported in 2-D TPD simulations, 
bringing it close to the results of ICF experiments.

\end{abstract}


\maketitle


Since the 1960s, the pursuit of inertial confinement fusion (ICF) driven by 
lasers has led to large-scale research on laser interaction with the plasmas 
of ICF targets \cite{Craxton2015}. 
Decades of laser--plasma interaction (LPI) research \cite{myatt14} have 
concentrated on several processes in laser-produced plasmas 
that can grow as parametric instabilities at high-enough laser intensities, 
namely stimulated Raman scattering (SRS), stimulated Brillouin scattering 
(SBS), and two-plasmon decay (TPD). 

Laser light can propagate in a plasma up to the critical density 
($n_\text{c}$) determined 
by the laser frequency. The region near quarter-critical density 
($\sfrac{1}{4}$ $n_\text{c}$) is a possible 
place for the interplay between SRS, SBS, and TPD as all three instabilities
can develop at that region. 
plasmons produced by SRS and TPD generate fast electrons that
can preheat the fusion fuel and 
degrade the performance of the ICF targets \cite{Craxton2015}, making 
LPI a concern in ICF experiments.
Several mechanisms of fast--electron acceleration have been studied before, 
namely staged acceleration \cite{yan09, yan12}, Langmuir 
cavitation \cite{vu12,vu12a}, and wave breaking \cite{coffey71}. 

In this Letter, LPI is studied using particle-in-cell (PIC) 
modeling \cite{Dawson83}, which can describe the interplay between different 
instabilities and the particle distributions including 
fast-electron generation. Usually, few hot electrons are found in 
the simulations at the linear stage of the TPD and SRS instabilities. 
The electron acceleration becomes effective after the instabilities 
saturate \cite{yan12}.

The TPD-related waves are mostly localized in the plane of 
polarization \cite{simon83}, which is defined by the incident laser wave 
vector (in the $x$ direction) and the laser electric field vector (in
the $y$ direction). 
The SRS sidescattering develops mostly outside of the polarization 
plane, and its scattered-light wave vector is almost perpendicular to the 
incident laser wave vector \cite{liu74,afeyan97c}. Scattered light waves 
can also propagate in the direction parallel or anti-parallel to the 
laser wave vector (forward- and backscattering, respectively) \cite{drake73}. 
A 2-D simulation in the polarization plane ($x$--$y$) or in the 
perpendicular plane ($x$--$z$) will be referred to as 
$p$ polarized (PP) or $s$ polarized (SP), respectively. 
Two-dimensional simulations can model only the interaction 
where either (in PP simulations) TPD or (in SP simulations) SRS 
dominates except for the high-frequency hybrid instability 
(HFHI) \cite{afeyan97b} case when the SRS 
scattered light propagates in the backward direction 
and the SRS-related and TPD-related waves are in the same ($x$--$y$) plane. 
The 3-D simulations are required to study the 
interaction including both TPD and SRS. In this Letter, the results 
of several 3-D simulations for different plasma parameters 
and incident laser profiles are presented and compared with the respective 
2-D simulations to illustrate that 
both TPD and SRS strongly influence the LPI near 
$\sfrac{1}{4}$ $n_\text{c}$. 
In the 3-D modeling including both TPD and SRS the fast-electron 
flux is reduced 
by up to an order of magnitude compared to 2-D TPD simulation results 
published before \cite{yan12}.

Here we describe in detail a 3-D simulation for the parameters relevant to 
ICF experiments \cite{seka09,Michel13}. 
A CH plasma is initialized with the electron temperature $T_\text{e}=2$ keV, 
and the temperatures for both ion species $T_\text{i}=1$ keV. 
The incident laser beam with intensity $I = 9\times10^{14}$ $\text{W/cm}^2$ 
propagates in the direction of density inhomogeneity ($x$). 
A linear density profile with the scale 
length $L= 100$ $\mu\text{m}$ is assumed at the initial time. 
The size of the simulation box is 
$21~\mu\text{m}\times8.4~\mu\text{m}\times6.7~\mu\text{m}$ 
modeling the density range from 0.21 $n_\text{c}$ to 0.26 
$n_\text{c}$. 

\begin{figure}[ht!]
\centering
\includegraphics[width=3.375in]{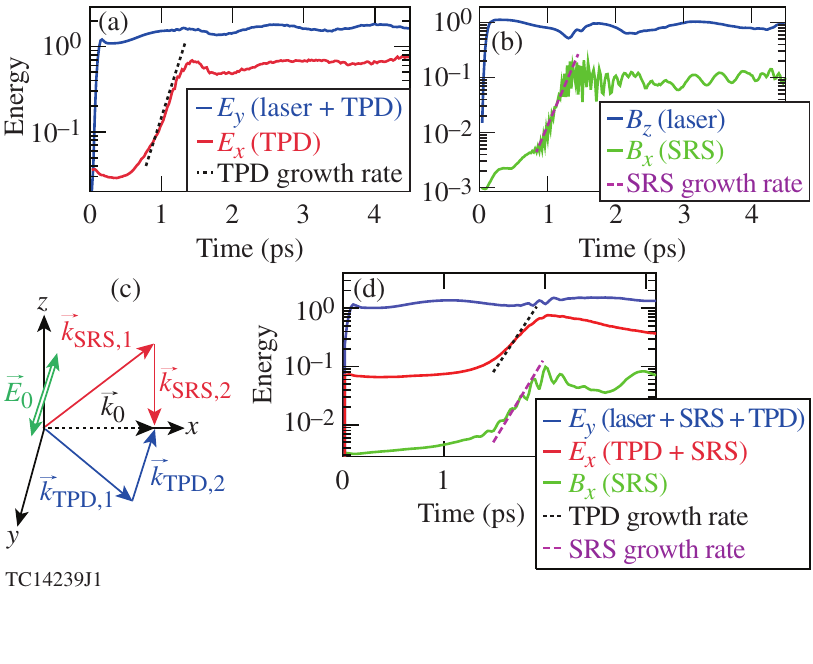}
\caption{The integrated energy of different field components in 
the simulation region as a function of time 
for the (a) 2-D $p$-polarized simulation,(b) 2-D $s$-polarized 
simulation, and (d) 3-D simulation. 
The wave-vector diagram for TPD and SRS is shown in (c).}\label{fig:figure1}
\end{figure}

Two 2-D simulations (PP and SP) with the same physical parameters were 
also performed. 
The TPD threshold parameter $\eta$ \cite{simon83} is 1.9 
($\eta=1$ at threshold), and the SRS backscattering threshold parameter 
$N$ \cite{drake73} is 0.5 ($N=0.26$ at SRS threshold) for these simulations.
The SRS sidescattering threshold \cite{liu74,afeyan97c} is close to the 
backscattering threshold for these parameters, and both absolute 
TPD and absolute SRS 
instabilities are expected to grow. The threshold of the convective 
SRS \cite{liu74} is not exceeded for the parameters described above. 
The time evolution of the energy of the field components in the 
simulation region is shown in Fig. \ref{fig:figure1}. 
The field energy is defined as the square of the electric- or 
magnetic-field amplitudes integrated over the simulation region 
normalized to the respective laser field energy 
at early time (when there are no instabilities). 
In the 2-D PP simulation [Fig. \ref{fig:figure1}(a)], the $E_x$ field 
contains most of the energy of 
the TPD plasmons with a larger wave vector. 
One can see that the field energy associated with the TPD instability 
stays at about the same level (close to 70\% of the energy of the 
incident laser electric field) after 2.5 ps, when one can assume that 
the saturation stage is reached. In the 2-D SP simulation, the energy 
of the $B_x$ field [Fig. \ref{fig:figure1}(b)] is used as an indicator 
for the level of SRS instability. The energy of the scattered light 
saturates at a level of 
about 8\% of the energy of the incident laser magnetic field. 

In the 3-D simulation, the diagram for TPD and SRS is shown in 
Fig. \ref{fig:figure1}(c), where the incident light ($\vec{k}_0$) 
decays into a plasmon ($\vec{k}_{\text{SRS},1}$) and a light wave 
($\vec{k}_{\text{SRS},2}$) in the case of SRS and into two 
plasmons ($\vec{k}_{\text{TPD},1}$ and $\vec{k}_{\text{TPD},2}$) 
in the case of TPD. 
The $E_x$ field energy [red line in Fig. \ref{fig:figure1}(d)] 
now includes the energy of the TPD plasmons 
and the SRS plasmons. The red line is overlaid with the dotted 
black line that represents the maximum TPD growth rate \cite{simon83}
($7.7\times10^{-4}~\omega_0$) minus the damping 
rate of plasma waves ($2.1\times10^{-4}~\omega_0$) measured in the 3-D 
simulation. The $E_x$ field saturates at a level of about 40\% of the 
laser field energy, which is much lower than the saturation level in 
the 2-D PP simulation. 
The green line in Fig. \ref{fig:figure1}(d) corresponds to the 
energy of the scattered light wave 
(propagating in the $z$ direction) from the SRS and is overlaid 
with the dashed purple line representing the maximum growth rate of 
the SRS \cite{liu74}($8.2\times10^{-4}~\omega_0$) 
minus the damping rate of the plasma waves. One can see from 
Fig. \ref{fig:figure1}(d) that the growth of the $B_x$ field 
energy in time is in reasonable agreement 
with the theoretical result \cite{liu74}. 
The $B_x$ field energy is under 10\% of the incident laser field 
energy after reaching its peak value, which is consistent with the 
2-D SP simulation result.

\begin{figure}[ht!]
\centering
\includegraphics[width=3.375in]{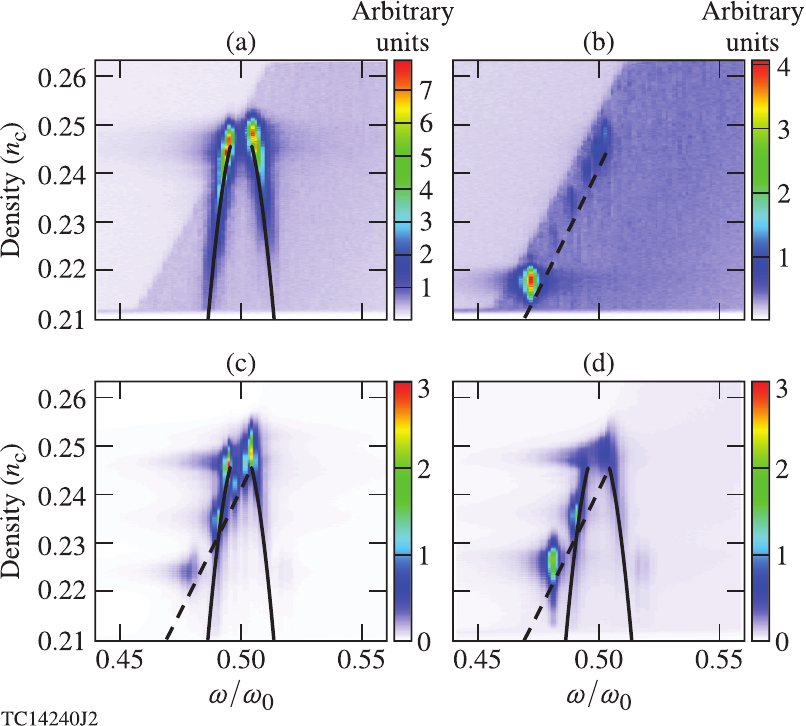}
\caption{(a) Plasma-wave spectra in the linear instability 
stage as a function of plasma density 
and the wave frequency normalized to laser frequency in the 
2-D PP simulation, (b) 2-D SP simulation,
and (c) 3-D simulation for modes with $0 \leq k_z/k_0 <0.2$ 
and (d) $0.2 \leq k_z/k_0 < 3$. 
The overlaid solid black lines and the dashed black lines 
represent the dispersion relations 
satisfying the matching conditions for TPD and 
SRS, respectively. }\label{fig:figure2}
\end{figure}

The spectra of plasma waves ($|\vec{E}_\text{L}|$) 
obtained at a time interval between 0.3 ps and 1.0 ps in the 2-D PP and SP 
simulations are plotted in Figs. \ref{fig:figure2}(a) and 
\ref{fig:figure2}(b), respectively. 
From the 3-D simulation, the spectra of plasma waves at a time interval 
between 1.3 ps and 2.0 ps are plotted 
in Fig. \ref{fig:figure2}(c) (close to $k_z=0$ plane, where TPD dominates) and 
in Fig. \ref{fig:figure2}(d) (far away from $k_z=0$ plane, where SRS 
dominates). One can see from Figs. \ref{fig:figure2}(c) and 
\ref{fig:figure2}(d) that TPD and SRS co-exist near 
$\sfrac{1}{4}$ $n_\text{c}$. 
The spectra of the unstable modes for TPD and SRS are close 
to the linear theory results 
(see overlaid lines in Fig. \ref{fig:figure2}). 

\begin{figure}[ht!]
\centering
\includegraphics[width=3.375in]{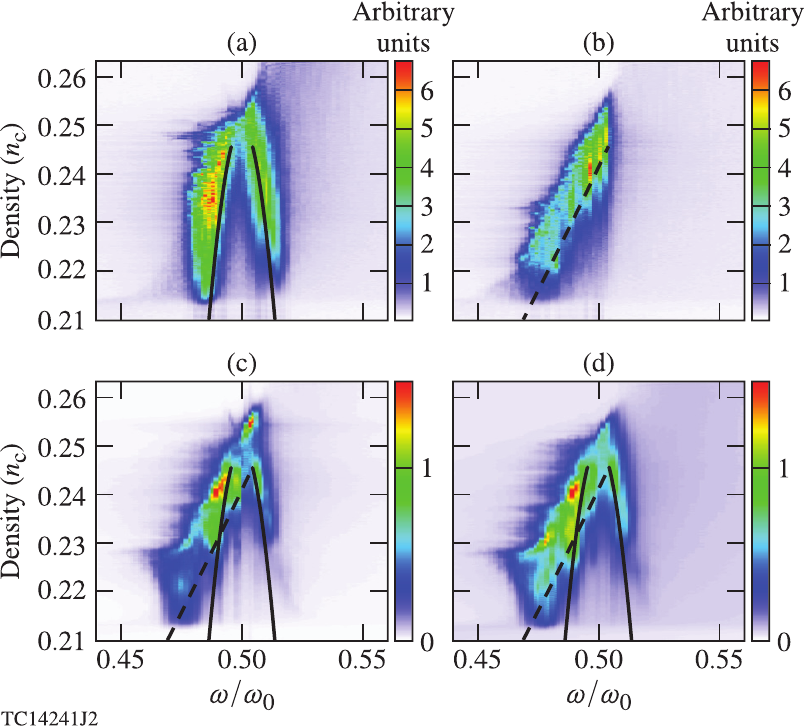}
\caption{Plasma-wave spectra in the saturation stage 
in the 2-D (from 3.3 ps to 4.1 ps)
and 3-D (from 2.3 ps to 3.1 ps) simulations 
as a function of plasma density and the wave frequency. 
Each panel displays the same quantity as in Fig. \ref{fig:figure2}.
}\label{fig:figure3}
\end{figure}

\begin{figure}[ht!]
\centering
\includegraphics[width=3.375in]{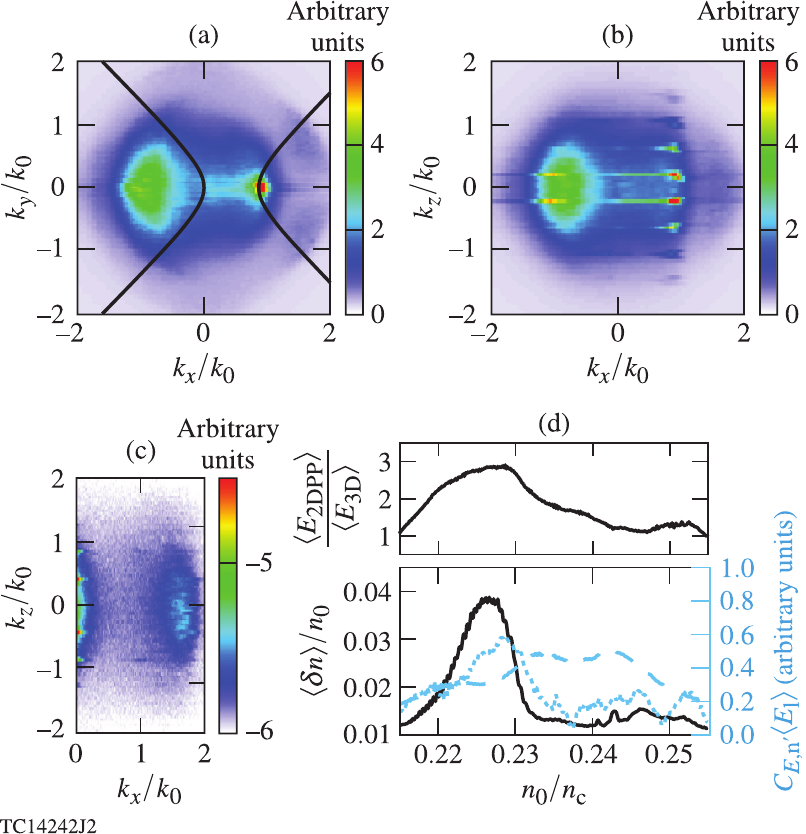}
\caption{(a) The spectrum of plasmons in the saturation stage of 3-D simulation 
at densities lower than $0.23~n_c$  plotted in the $k_x$--$k_y$ plane 
and (b) the $k_x$--$k_z$ plane.
(c) The spectrum of ion density fluctuation plotted in 
the $k_x$--$k_z$ plane on a logarithmic scale. 
(d) Lower panel: Ion density fluctuations RMS (root-mean-square
average over the transverse direction and time) normalized to 
background density (black solid line), 
longitudinal electric field RMS (blue dashed line) 
and caviton correlator $C_{E,n}$ (blue dotted line). 
Upper panel: the ratio of the electric field amplitude 
of the TPD plasmons with larger wave vector between 
2-D PP and 3D simulations.} \label{fig:figure4}
\end{figure}

As the instability evolves from the linear stage to the saturation stage, 
the frequency spectra shown in Fig. \ref{fig:figure2} evolve 
into the spectra shown in Fig. \ref{fig:figure3}. 
One can see that 
the spectra in all these simulations are broader in the 
saturation stage compared to the linear stage. 
The density in Fig. \ref{fig:figure3} is calculated using 
the initial density profile.  
Compared to the 2-D PP simulation [Fig. \ref{fig:figure3}(a)], 
the TPD is much weaker at densities 
lower than 0.23 $n_\text{c}$ in the 3-D simulation [Fig. \ref{fig:figure3}(c)]. 
The weakening of the TPD modes at these densities is also 
illustrated in Fig. \ref{fig:figure4}(a) [and Fig. \ref{fig:figure4}(b)], 
where the spectrum of plasma waves at densities below 0.23 
$n_\text{c}$ in the saturation stage is integrated over $k_z$ (and $k_y$). 
There are no prominent modes along the TPD hyperbola 
\cite{meyer93} [black solid line in Fig. \ref{fig:figure4}(a)] 
at $k_x > k_0$, which corresponds to the TPD daughter waves 
with larger wave vectors. 
Two types of low-frequency density fluctuations 
are identified in our simulations [see Fig. \ref{fig:figure4}(c)]. 
One type are the ion acoustic waves driven by 
the Langmuir-decay instability (LDI) \cite{DuBois67,DuBois96} 
and the other type are driven with the beating of the same-frequency 
daughter waves of SRS and TPD. 
The LDI modes form a
broad feature at $k_x \approx 1.7~k_0$ (about $2\times$ the
laser wave vector in plasma) in the spectrum of the ion
density fluctuations shown in Fig. \ref{fig:figure4}(c).
The beating of the SRS plasmons with wave vector
$(k_x,~k_y,~k_z) = (0.87~k_0,~0,~\pm 0.2~k_0)$ creates density
perturbations with wave vector $(k_x,~k_z) = (0,~\pm 0.4~k_0)$. The
coupling between SRS plasmons and density perturbations generates higher-order
modes in the field at $k_z=\pm (0.2 + m 0.4~k_0)$, [$m=1,~2,~3~...$,
as shown in Fig. \ref{fig:figure4}(b)] and in the density perturbation
at $(k_x,~k_z)=[0,~\pm (0.4 + m 0.4~k_0)]$ [see Fig. \ref{fig:figure4}(c)].

Although SRS and TPD grow independently in the linear stage, 
in the nonlinear stage they interact through 
low-frequency density perturbations. 
TPD growth starts from the region near $\sfrac{1}{4}$ $n_\text{c}$ 
and spreads to lower densities \cite{yan12} before being 
saturated by ion density perturbations. 
Compared to TPD saturation in 2-D (without SRS) ion 
density perturbations are much larger 
in 3-D (with both SRS and TPD), especially near the plasma 
region where the frequencies 
of TPD and SRS plasmons are close. In this region where the dispersion lines 
for TPD and SRS plasmons intersect [near 0.23 $n_\text{c}$ in our simulations, 
see Fig. \ref{fig:figure3}(c) and \ref{fig:figure3}(d)]
multiple pairs of SRS and TPD daughter waves have close frequencies 
and can drive ion density perturbations through the ponderomotive force 
to much higher levels compared to other density regions 
[see the black line in Fig. \ref{fig:figure4}(d)]. 
The growth of TPD plasmons at densities below 0.23 $n_\text{c}$ is disrupted by 
these enhanced ion density perturbations, as illustrated by a decrease  
in the level of TPD-driven plasmons below 
0.23 $n_\text{c}$ in Fig. \ref{fig:figure3}(c).

The correlation between the 
local plasmon intensity $|E_L|^2$ and 
the density fluctuations $\delta n$ is captured using the 
caviton correlator \cite{vu12} $ C_{E,n} =  \langle - \delta n 
|E_L|^2 \rangle / (\langle (\delta n)^2 \rangle^{1/2}
\langle |E_L|^2 \rangle)$. 
As shown in the lower panel of Fig. \ref{fig:figure4}(d), 
the plasma waves and the density fluctuations are
weakly correlated between 0.255 $n_\text{c}$ and 0.235 
$n_\text{c}$ : $C_{E,n} = 0.1-0.2$ 
in spite of a significant level of plasmons in this density range.
At densities close to 0.23 $n_\text{c}$, lower 
panel of Fig. \ref{fig:figure4}(d) 
shows the increase not only in the plasmon intensity and density fluctuations, 
but also in the correlation between them with $C_{E,n}$ reaching up to 0.6. 
The large caviton correlator indicates that the plasma waves are strongest 
in areas where density is depleted. The ponderomotive force of 
multiple pairs of SRS and TPD daughter waves with close frequencies 
is responsible for driving the enhanced density perturbations. 
The nonlinear coupling of TPD and SRS through ion perturbations leads to 
a lower TPD saturation level in the 3-D simulation 
compared to the 2-D PP simulation, which is illustrated in the upper 
panel of Fig. \ref{fig:figure4}(d).

The fast-electron flux is defined as 
the energy flux carried by electrons with kinetic energy above 55 keV 
leaving the simulation box minus the energy flux carried 
by the thermal electrons 
injected into the simulation region from the thermal 
boundaries (in the $x$ direction).
Information about the hot electrons is collected during the 
saturation stage in each simulation
for 0.5 ps.
In the 3-D simulation, the fast-electron flux associated 
with the forward- and backward-going hot electrons was
found to be 1.7\% and 0.8\%, respectively.
The plasma-wave spectrum in the 3-D simulation corresponds 
to a smaller $k$-space domain than 
the spectrum in the 2-D PP simulation, which 
makes the staged acceleration 
mechanism less efficient in 3-D than in 2-D and
explains a smaller number of hot electrons in the 3-D simulation 
compared to the 2-D PP simulation (6.6\% and 3.4\% in the 
forward and backward direction, respectively).
The influence of wave-breaking on the fast-electron generation 
is small as the maximal electric field amplitude (0.04 $m_e \omega_0 c / e$) is  
below the wave-break limit(0.1 $m_e \omega_0 c / e$) \cite{coffey71}.

The nonlinear regime including both TPD and SRS  
is also observed in simulations 
with the speckled laser beam \cite{kato84,skupsky89} and 
electron--ion collision effects included. 
The speckled laser beam is modeled by a single speckle in the simulation region 
that mirrors itself in the transverse direction because 
of the periodic boundary conditions. 
A series of simulations has been performed to study how the 
speckles affect the generation of hot electrons. 
All parameters are the same as the simulations described previously 
except for the temperatures of electrons and ions being 1.5 times higher. 
The peak intensities in the laser speckles are 
$1.8 \times 10^{15}~\text{W/cm}^2$ (twice of the average intensities). 
A collision package (CP) is available for the PIC code 
OSIRIS \cite{fonseca02}. 
The main physics processes are observed to be the same 
in simulations with plane-wave beams and speckled beams.

The fast-electron flux values in simulations are listed in 
Table \ref{tab:eflux} for 
different incident laser beams as well as with CP turned on and off. 
By comparing the left and right columns of Table \ref{tab:eflux}, one 
can see that adding collisions can reduce the fast-electron flux 
by about 50\% and in the case of plane-wave 2-D PP simulation by 
almost 70\%. Also note that the reduction 
of the fast-electron flux caused by collisions affects both 
the forward-going electrons and backward-going electrons since 
the collisional damping rate affects all the plasma waves. 
The fast-electron flux generated in the 2-D SP simulations 
is much smaller than the fast-electron flux generated in 
the 2-D PP simulations, 
which indicates that the plasma waves driven by TPD are the 
main source of the electron acceleration. 

The hot electron fraction observed in  
the ICF experiments on the OMEGA laser system 
does not exceed few percent \cite{Michel13}. 
At the same time, in the previous PIC simulations 
of TPD in 2-D the hot electron fraction 
was close to an order of magnitude larger than in the experiments. 
The 3-D PIC simulations presented in this Letter for the first time 
produce the results for the hot electron fraction
that are close to the experimental levels.

\begin{table}
\begin{center}
\begin{tabular}{ c | c | c }
\hline \hline
   Fast-electron flux &  \multicolumn{2}{c}{Forward/Backward} \\
\hline
 Collision package & On & Off \\
\hline
Plane wave 2-D PP        & $1.6\%/1.3\%$        &  $5.5\%/3.8\%$  \\
\hline
Plane wave 2-D SP        & $(<0.1\%)/0.2\%$   &  $(<0.1\%)/0.5\%$  \\
\hline
Speckle 2-D PP             & $6.8\%/1.7\%$        & $9.4\%/3.8\%$  \\
\hline
Speckle 2-D SP             & $(<0.1\%)/0.3\%$   & $(<0.1\%)/0.7\%$ \\
\hline
Speckle 3-D                 & $0.4\%/0.3\%$        &  $0.8\%/0.5\%$\\
 \hline\hline
\end{tabular}
\end{center}
\caption{Fast-electron flux normalized to the incident laser energy flux.}
\label{tab:eflux}
\end{table}

laser--plasma interaction near $\sfrac{1}{4}$ $n_\text{c}$ 
determines the generation of fast electrons 
that are crucial for the performance of ICF targets.
The fast-electron flux in simulations is found 
to be closely related to the plasma-wave spectra. 
The TPD-driven plasma waves with large wave vectors 
are very important for accelerating electrons. 
At the same time, the SRS-driven plasma waves are less 
effective in accelerating electrons.
Therefore the modeling including 
the nonlinear coupling of TPD and SRS 
in 3-D is the only way 
to correctly describe the generation of fast electrons in laser-driven ICF.

Our 3-D PIC simulations have shown  
the large decrease (up to an order of magnitude) in the fast-electron flux 
compared to 2-D TPD modeling. 
The reason is the nonlinear coupling between 
SRS and TPD which is especially pronounced 
at densities lower and around 0.23 $n_\text{c}$.  
In this region plasma waves and growing density perturbations 
are localized in same areas as illustrated by the caviton correlator. 
Enhanced density perturbations detune and weaken the TPD-driven plasmons effective 
in the fast electron generation. 
In addition to the TPD suppresion, the plasma wave spectra 
in 3-D simulations are much more narrow 
compared to the spectra in 2-D TPD modeling.
To conclude, 3-D PIC simulations presented in this Letter 
fully model the laser-plasma interaction near 
$\sfrac{1}{4}$ $n_\text{c}$ including SRS and TPD, 
and obtain the fast electron fraction level close to experimental results, 
resolving the large discrepancy between ICF experiments 
and PIC simulations that existed for many years before.

\begin{acknowledgments}
This work was supported by the Department of Energy 
National Nuclear Security Administration under Award Number DE-NA0001944, 
the University of Rochester, and the New York State 
Energy Research and Development Authority. 
We also acknowledge the support by the DOE under grant 
No. DE-SC0012316, 
and by the NSF under grant No. PHY-1314734. This report 
was prepared as an account of work 
sponsored by an agency of the U.S. Government. Neither the U.S. 
Government nor any agency 
thereof, nor any of their employees, makes any warranty, express 
or implied, or assumes any 
legal liability or responsibility for the accuracy, completeness, 
or usefulness of any 
information, apparatus, product, or process disclosed, or 
represents that its use would not 
infringe privately owned rights. Reference herein to any 
specific commercial product, process, 
or service by trade name, trademark, manufacturer, or 
otherwise does not necessarily 
constitute or imply its endorsement, recommendation, or 
favoring by the U.S. Government 
or any agency thereof. The views and opinions of authors 
expressed herein do not necessarily 
state or reflect those of the U.S. Government or any agency thereof. 
\end{acknowledgments}

\end{document}